\documentclass{article}
\usepackage{spconf,amsmath,graphicx,siunitx,microtype,amsfonts}
\usepackage[numbers]{natbib}
\usepackage{array}
\usepackage{booktabs}

\ninept

\title{DIVERSE AUDIO CAPTIONING VIA ADVERSARIAL TRAINING}
\name{Xinhao Mei, Xubo Liu, Jianyuan Sun, Mark D. Plumbley, Wenwu Wang}
\address{Centre for Vision, Speech and Signal Processing (CVSSP)\\
        University of Surrey, UK\\
        \{x.mei, xubo.liu, jianyuan.sun, m.plumbley, w.wang\}@surrey.ac.uk\\ }
\begin{document}
%
\maketitle
\begin{abstract}
Audio captioning aims at generating natural language descriptions for audio clips automatically. 
Existing audio captioning models have shown promising improvement in recent years.
However, these models are mostly trained via maximum likelihood estimation (MLE), which tends to make captions generic, simple and deterministic. 
As different people may describe an audio clip from different aspects using distinct words and grammars, we argue that an audio captioning system should have the ability to generate diverse captions for a fixed audio clip and across similar audio clips. To address this problem, we propose an adversarial training framework for audio captioning based on a conditional generative adversarial network (C-GAN), which aims at improving the naturalness and diversity of generated captions. Unlike processing data of continuous values in a classical GAN, a sentence is composed of discrete tokens and the discrete sampling process is non-differentiable. To address this issue, policy gradient, a reinforcement learning technique, is used to back-propagate the reward to the generator. The results show that our proposed model can generate more diverse captions, as compared to state-of-the-art methods.
\end{abstract}
\begin{keywords}
Audio captioning, GANs, deep learning, cross-modal task, reinforcement learning
\end{keywords}
\section{Introduction}
\label{sec:intro}

Generating natural language descriptions for audio clips, known as automated audio captioning (AAC), has received increasing attention recently as it can be widely used in applications such as assisting for the hearing-impaired to understand environmental sounds, summarizing multimedia content, and analyzing sounds for security surveillance. Compared with other audio-related tasks (i.e. audio tagging and sound event detection \cite{kong2020panns}), audio captioning is a more challenging cross-modal task which needs to capture the spatial-temporal relationships between the presented audio events and express them in natural language.

Benefiting from the development of deep learning and the release of high-quality datasets, audio captioning systems have achieved great improvement in recent years \cite{kim2019audiocaps, drossos2020clotho}. Most existing audio captioning models are trained via maximum likelihood estimation (MLE), which encourages the use of high frequent $n$-grams and common expressions occurred in the training set. As a result, the generated captions can faithfully describe the content of the audio clip and achieve high score on the $n$-gram based metrics such as BLEU \cite{papineni2002bleu} and CIDE$_r$ \cite{vedantam2015cider}. However, captions generated by models trained via MLE are usually deterministic (generating a fixed caption for a given audio clip), simple (using simple and common words) and generic (generating same caption for similar audio clips). 

In practice, given an audio clip, people may focus on different parts and tend to describe the content from various aspects using distinct words, phrases and grammars. As a result, the generated captions may be of great diversity. For example, in the audio captioning datasets \cite{drossos2020clotho}, each audio clip comes with several diverse, human-annotated captions. Unlike accuracy, diversity is usually neglected but is an important property that an audio captioning system should possess. In this work, we are motivated to develop an audio captioning system that can generate diverse captions either for a fixed audio clip or similar audio clips.

\begin{figure*}[!t]
  \centering
  \includegraphics[width=\linewidth]{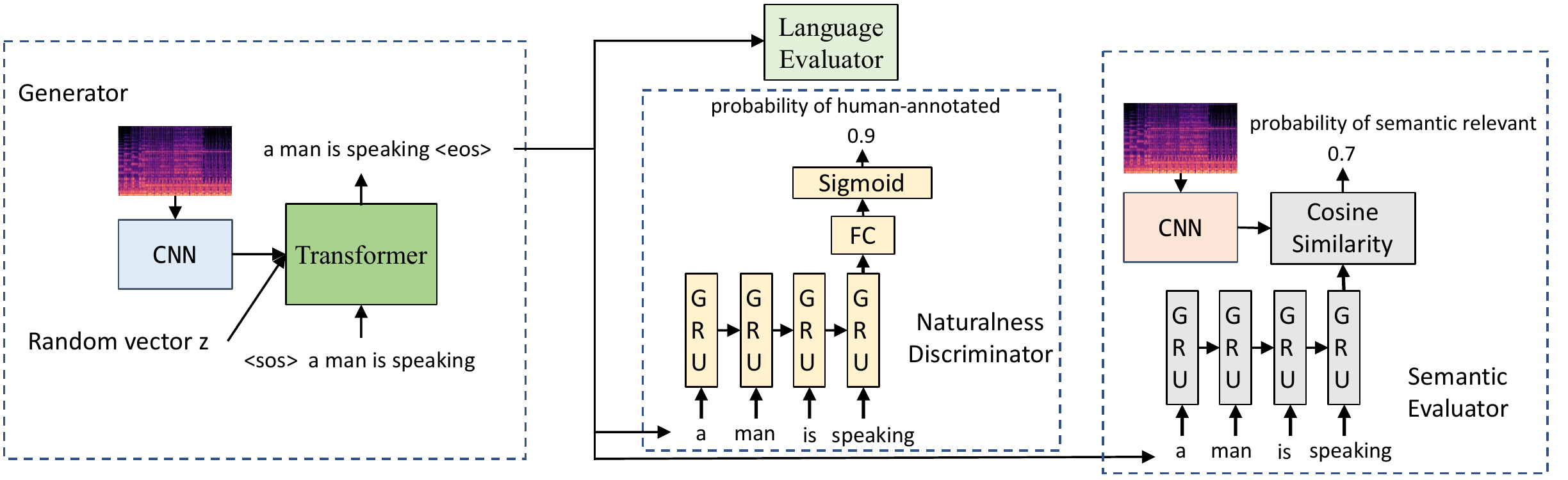}
  \caption{Overview of the adversarial training framework, where the generator $G_{\theta}$ aims at generating captions which can confuse the discriminator $D_{\phi}$, while $D_{\phi}$ aims at correctly classifying human-annotated and machine-generated captions. The language evaluator evaluates captions using a conventional evaluation metric CIDE$_r$. The semantic evaluator evaluates whether the generated captions are faithful to given audio clips.}
  \label{fig:framework}
\end{figure*}

Instead of training the model using MLE, we propose an adversarial training framework based on conditional generative adversarial network (C-GAN) \cite{goodfellow2014generative}, which has been successfully used to generate natural images. Our proposed model is composed of a generator, a discriminator and two evaluators. The generator is trained to generate natural captions while the discriminator is trained to distinguish between the machine-generated and human-annotated captions. These two modules compete with each other and are trained in an alternating manner. To ensure the accuracy and diversity simultaneously, a language evaluator and a semantic evaluator are introduced to evaluate the generated captions, where the language evaluator evaluates captions using one conventional evaluation metric CIDE$_r$ and the semantic evaluator evaluates whether the generated captions are faithful to the audio clips or not. 
As a caption is composed of discrete tokens, it is not feasible to make slight changes to the discrete output value with respect to the gradients back-propagated from the discriminator.
Inspired by SeqGAN \cite{yu2017seqgan}, we tackle this problem by policy gradient \cite{sutton2000policy}, a reinforcement learning technique. The scores from the discriminator and evaluators are regarded as a reward which the generator is trained to maximize.

To our knowledge, this is the first work to explore the diversity issue in an audio captioning system in the literature. The results show that our method can generate diverse and accurate captions, as compared to other state-of-the-art methods.

\section{Related works}
\label{sec:related_works}

Audio captioning is generally regarded as a cross-modal text generation problem. Inspired by the great success of encoder-decoder paradigm in natural language processing (NLP) tasks, most existing audio captioning models use an encoder to extract audio features, and a decoder to generate captions based on the extracted features \cite{kim2019audiocaps, drossos2017automated}. CNN-RNN \cite{xu2021investigating} and CNN-Transformer \cite{chen2020audio} are the two dominant architectures which achieve state-of-the-art performance, while Transformer-only network also shows competitive performance \cite{mei2021audio}. To avoid the indeterminacy of word selection, keywords estimation was introduced as auxiliary information \cite{koizumi2020keywordtrans, ye2021_t6}. Koizumi et al \cite{koizumi2020audio} adopted a large pre-trained language model GPT-2 and audio-based similar caption retrieval to guide the caption generation. Liu et al \cite{liu2021cl4ac} proposed a contrastive loss to improve the quality of latent representation and the alignment between audio and texts. 

Audio captioning models are generally trained via MLE with `teacher-forcing' strategy, where the model is trained to maximize the conditional log-likelihood of ground-truth word given audio features and previous ground-truth words using cross-entropy (CE) loss. The `teacher-forcing' training strategy suffers from the exposure bias problem which leads to error accumulation during inference \cite{bengio2015scheduled}, since the model generates words based on ground-truth words during training, but it generates words based on previous generated words at test time. Reinforcement learning is introduced to address this exposure bias issue and to directly optimize the evaluation metrics, such as CIDE$_r$ \cite{ mei2021encoder}. 

The performance of an audio captioning system is usually evaluated by the $n$-gram based metrics, such as BLEU \cite{papineni2002bleu}, ROUGE$_l$ \cite{lin2004rouge} and CIDE$_r$ \cite{vedantam2015cider}. These metrics will give higher scores to sentences having frequent $n$-grams occurred in the training set rather than those using variant words or grammars. Another popular metric is SPICE \cite{anderson2016spice}, which is based on scene graphs. Existing audio captioning systems trained via MLE are capable of generating captions which can accurately describe the content of the audio clips, and achieving high scores regarding to above evaluation metrics. However, the models trained via MLE tend to generate generic, simple and rigid captions, even if each audio clip has multiple and diverse ground-truth captions in the training set. The use of reinforcement learning also reduces distinctiveness in the captions generated \cite{mei2021encoder}.

To our knowledge, almost all work in audio captioning have focused on improving the accuracy of the captions and the scores of the evaluation metrics so far, none of them pay attention to the diversity of the generated captions. Inspired by some related works in image captioning \cite{dai2017towards, li2018generating, shetty2017speaking}, we aim to improve the diversity of the generated captions via adversarial training.


\section{PROPOSED FRAMEWORK}
\label{sec:framework}

The proposed framework consists of a caption generator $G_{\theta}$, a discriminator $D_{\phi}$ and two evaluators, where $G_{\theta}$ and $D_{\phi}$ make up a conditional generative adversarial network (C-GAN) and are trained alternatively to compete with each other, $\theta$ and $\phi$ are the parameters of the generator and discriminator respectively. The two evaluators include a language evaluator which evaluates the caption with the performance metric, CIDE$_r$, and a semantic evaluator which evaluates whether the caption is faithful to the given audio clip or not. These two evaluators are fixed during the adversarial training process. The diagram of the proposed adversarial training framework is shown in Fig.~\ref{fig:framework}.

\begin{table*}[!t]
\centering
\begin{tabular}[\linewidth]{c c c c c | c c c c} 
 \hline
  & $\lambda$ & BLEU$_{4}$ ($\uparrow$) & CIDE$_r$ ($\uparrow$) & SPIDE$_r$ ($\uparrow$) & vocab size ($\uparrow$) & mBLEU$_{4}$ ($\downarrow$) & div-1 ($\uparrow$) & div-2 ($\uparrow$) \\  
 \hline
 MLE & & \textbf{0.162} & \textbf{0.393} & \textbf{0.256} & 746 & 0.793 & 0.310 & 0.361 \\
 C-GAN & 1.0 & 0.122  & 0.295 & 0.199 & \textbf{818} & \textbf{0.615} & \textbf{0.335} & \textbf{0.442} \\
 C-GAN & 0.7 & 0.138 & 0.326 & 0.216 & 776 & 0.661 & 0.326 & 0.413 \\
 C-GAN & 0.5 & 0.148 & 0.347 & 0.229 & 763 & 0.647 & 0.320 & 0.413  \\
 C-GAN & 0.3 & 0.157 & 0.366 & 0.240 & 560 & 0.703 & 0.291 & 0.371 \\
 C-GAN & 0.0 & 0.155 & 0.362 & 0.236 & 218 & 0.792 & 0.225 & 0.345  \\
 \hline
\end{tabular}
\caption{Performance of the CNN-Transformer network trained via MLE and our proposed C-GAN framework. 
BLEU$_4$, CIDE$_r$ and SPIDE$_r$ are conventional evaluation metrics. Vocabulary size, mBLEU$_4$, and div-$n$ are the diversity metrics.}
\label{table:tab_results} 
\end{table*}

\subsection{Generator}
\label{ssec:generator}
The state-of-the-art CNN-Transformer network is used as the caption generator \cite{mei2021encoder}. It is worth noting that the proposed adversarial training framework is agnostic to the architecture of the caption generator, which is also applicable to other types of caption generator (e.g. RNN-RNN \cite{drossos2017automated}, CNN-RNN \cite{xu2021investigating}).

A 10-layer pre-trained audio neural network (PANN) is used as the audio encoder to extract audio features \cite{kong2020panns}. Given an audio clip $x$, the encoder takes the log mel-spectrogram as input and produces the audio feature vectors $f(x)$, which are then concatenated with a random vector $z$ sampled from a normal distribution and fed into the Transformer decoder. In each time step, the word can be sampled as follows:
\begin{equation}
  \label{eqn:words_sample}
  w_{t} \sim \pi _{\theta}(w_t | x, z, w_{1:t-1})
\end{equation}
where $\pi_{\theta}$ is the conditional distribution over the vocabulary, and $w_t$ is the word sampled at time step $t$. The decoder can generate diverse captions given the same audio clip and different random vectors. 

During the training, $G_{\theta}$ observes three scores from $D_{\phi}$ and the two evaluators, based on different criteria, and is trained to maximize these scores. 
However, unlike in classical GANs where the data to be processed are of continuous values, the captions are composed of discrete tokens and the discrete sampling process is non-differentiable. As a result, the scores from discriminator and evaluators cannot be optimized through back-propagation directly. To address this issue, policy gradient \cite{yu2017seqgan}, which is a reinforcement learning (RL) technique, is adopted to address this problem. 
In a RL setting, the decoder acts as an agent which can interact with an environment (words in the vocabulary and the audio features). The generation of each word is an action of the agent governed by a policy $\pi_{\theta}$ defined by the parameters of $G_{\theta}$. Upon generating a complete sentence, the agent can observe a reward from $D_{\phi}$ and evaluators. The objective of the agent is to take a sequence of actions to maximize the final expected reward, which we formulated as minimizing the negative of the final expected reward:
\begin{equation}
  \label{eqn:rl_g}
  \mathcal L_{\rm G}(\theta)= -\mathbb E_{w \sim \pi_\theta}[r(w)]
\end{equation}
where $w=(w_1,...,w_T)$ is a generated caption from $G_{\theta}$. The reward $r(w)$ is calculated as:
\begin{equation}
  \label{eqn:reward}
  r(w) = \lambda (n + s) + (1 - \lambda) \cdot c
\end{equation}
where $n$ and $s$ are the scores output by the $D_{\phi}$ and semantic evaluator, respectively, $c$ is the CIDE$_r$ score of the caption from the language evaluator and $\lambda$ is a hyper-parameter with a value between \num{0} and \num{1}. We group $n$ and $s$ as they ensure semantic relevance and naturalness together. When $\lambda$ equals 0, the system degenerates to a conventional RL method which optimizes the evaluation metric directly. We employ the self-critical sequence training (SCST) method \cite{rennie2017scst} to calculate the gradient of the objective function $\nabla_\theta \mathcal L_{\rm G}(\theta)$, which can be formulated as:
\begin{equation}
  \label{eqn:rl_gradient}
  \nabla_\theta \mathcal L_{\rm G}(\theta) \approx - \sum_{t=1}^T (r(w) - r(\hat{w}))\nabla_\theta \log \pi_\theta (w_t|x, z, w_{1:t-1})
\end{equation} 
where $r(\hat{w})$ is the reward of the caption sampled by the current model using greedy search and used as a baseline to normalize $r(w)$. As a result, only samples from the model having a higher reward than those greedy-decoding samples are given positive weights.

\subsection{Discriminator and evaluators}
\label{ssec:discriminator}

A  single-layer gated recurrent unit (GRU) \cite{chung2014gru} is used as the discriminator $D_{\phi}$, which takes a caption as input and outputs a probability $n \in [0, 1]$ that indicates whether the given caption is human-annotated or machine-generated. The training objective of $D_{\phi}$ can be formulated as:
\begin{equation}
  \label{eqn:eqn_dis}
  \mathcal{L}_{\rm D}(\phi) = \mathbb E_{c \sim \mathcal{C}_r}\log{D_{\phi}(c)} + \mathbb E_{c \sim \mathcal{C}_g}\log{ [1 - D_{\phi}(c) ]} 
\end{equation}
where $C_r$ is the human-annotated caption set and $C_g$ is the caption set generated by $G_{\phi}$. Here $D_{\phi}$ ensures the naturalness of the caption.

The semantic evaluator is a neural network composed of a CNN and a GRU, pre-trained using ground-truth captions and audio clips. The CNN extracts an audio embedding from a given audio clip, and the GRU extracts an language embedding for a given caption. The semantic evaluator is trained to return a high score of cosine similarity for paired audio and caption and a low score for unpaired ones. The semantic evaluator ensures the semantic relevance of the generated caption with the given audio clip. The language evaluator is based on the metric CIDE$_r$ and it ensures the captions achieve high scores under the objective evaluation metrics.

\begin{table*}[!t]
\centering
\begin{tabular}[\linewidth]{c c c c c | c c c c} 
 \hline
  &  BLEU$_{4}$ ($\uparrow$) & CIDE$_r$ ($\uparrow$) & SPIDE$_r$ ($\uparrow$) & vocab size ($\uparrow$) & mBLEU$_{4}$ ($\downarrow$) & div-1 ($\uparrow$) & div-2 ($\uparrow$) \\  

 \hline
 C-GAN\_ND & 0.047 & 0.071 & 0.054 & 642 & 0.156 & 0.591 & 0.824 \\
 C-GAN\_SE & 0.000 & 0.012 & 0.025 & 224 & 0.715 & 0.096 & 0.269 \\
 C-GAN\_LE & 0.155 & 0.362 & 0.236 & 218 & 0.792 & 0.225 & 0.345 \\
 \hline
\end{tabular}
\caption{Ablation studies of the discriminator and two evaluators. ND: naturalness discriminator, SE: semantic evaluator, LE: language evaluator with the CIDE$_r$ metric.}
\label{table:ablation_results} 
\end{table*}

\begin{table*}[ht]
\centering
\begin{tabular}[\linewidth]{c | c | c} 
 \hline
   & C-GAN & MLE \\ 
 \hline
 & a loud buzzing occurs and then starts again & a jackhammer is being used to start a stop \\
 & a jackhammer is being started and then slows down & a jackhammer is being used to get a stop \\
 Example 1 & a loud drilling occurs and then slows down & a jackhammer is being used to start up and down \\
 & an old engine is running and then loud squealing & a jackhammer is being used to start up \\
 & a loud drilling occurs and then starts again & a jackhammer is being used to start a halt \\
 \hline
 & the rain falls in the background as the rain falls in the background & it is raining and the wind is blowing \\
 & heavy rain is falling and the wind is blowing & it is raining and the rain is falling \\
 Example 2 & rain is pouring down on a windy day & it is raining and the rain is pouring down \\
 & the rain falls and thunder rolls in the background & it is raining and the rain is hitting the ground \\
 & the rain is pouring down on a windy day &it is raining and the rain is falling down \\
 \hline
\end{tabular}

\caption{Examples of the generated captions by proposed C-GAN model and MLE model. }
\label{table:tab_caps} 
\end{table*}

\section{EXPERIMENTS}
\label{sec:exp}

\subsection{Dataset}
\label{ssec:dataset}
All the experiments are conducted on the Clotho dataset \cite{drossos2020clotho} which contains \num{3839}, \num{1045} and \num{1045} audio clips in training, validation and evaluation set, respectively. Each audio clip has five diverse human-annotated captions. In line with our previous work, training set and validation set are merged together to give a new training set containing \num{4884} audio clips.


\subsection{Implementation details}
\label{ssec:impl_detail}
All audio clips in the dataset are converted to \num{44.1} KHz. We use \num{64}-dimensional log mel-spectrograms extracted by a \num{1024}-point Hanning window with a hop size of \num{512} as the input audio features. To process the captions in the dataset, we convert all the characters to lower case and remove the punctuation. Two special tokens ``\texttt{\textless sos\textgreater}'' and ``\texttt{\textless eos\textgreater}'' are padded to indicate the start and end of each caption. After these pre-processing steps, we get a vocabulary of size \num{4637}. 

The generator $G_{\theta}$ is pre-trained using MLE for \num{25} epochs following the setting in \cite{mei2021encoder}, and the discriminator $D_{\phi}$ is pre-trained for \num{5} epochs thus it can provide correct guidance in the initial training stage of GAN. The semantic evaluator is pre-trained for \num{25} epochs and is frozen during the adversarial training process. In the adversarial training stage, $G_{\theta}$ and $D_{\phi}$ are trained jointly for \num{30} epochs, in which one step of discriminator update is followed by one step of generator update. The batch size and learning rate is set to \num{32} and \num{0.0001} respectively. During the test stage, beam search with a beam size of \num{5} is used to sample the captions. One important hyper-parameter needed to be determined is $\lambda$. We test empirically with different values of $\lambda$ to find a proper balance between the rewards.

\subsection{Evaluation}
\label{ssec:eval_metrics}
We evaluate the system performance using $n$-gram based conventional evaluation metrics (BLEU$_n$, CIDE$_r$ and SPIDE$_r$) and diversity metrics (vocabulary size, mBLEU$_n$ and div-$n$). BLEU$_n$ and CIDE$_r$ are calculated based on $n$-gram matches between predicted caption and ground truths. SPIDE$_r$ is the average of CIDE$_r$ and SPICE \cite{anderson2016spice}, where CIDE$_r$ ensure syntactic fluency and SPICE ensures semantic relevance.
The vocabulary size is the number of unique words used in generated captions for the evaluation set. mBLEU$_n$ compares $n$-gram matches between one of the generated caption and the remaining generated captions for an audio clip, thus lower mBLEU$_n$ score means more diversity. Div-$n$ is calculated by comparing the number of distinct $n$-grams to the total number of words generated by a set of captions given an audio clip.

Table \ref{table:tab_results} shows the performance of the CNN-Transformer model trained via MLE and the proposed C-GAN framework. For the MLE model, beam search with a beam size of \num{5} is used to sample \num{5} distinct captions per audio clip, which are deterministic (generating the same captions when the same audio clip is given). For the proposed model, we sample \num{5} captions for each audio clip. For the $n$-gram based conventional metrics, the MLE model performs better than that trained via C-GAN. This is not surprising, as MLE training will encourage the use of high frequent $n$-grams occurred in references and these metrics mainly focus on $n$-gram matches with references. For the diversity metrics, the results clearly suggest that our proposed framework can generate more diverse captions, especially when $\lambda$ is close to \num{1.0}. When $\lambda$ is small, the reward from language evaluator contributes most, and optimizing CIDE$_r$ directly reduces the distinctness of the captions but improves the conventional metrics. Overall, considering these two types of metrics together, the captions generated by the MLE model may have better quality but are generic and lack of diversity. The captions generated by our proposed C-GAN framework are more diverse, but the reinforcement learning may introduce some mistakes in terms of the grammar, thus give a lower score on the $n$-gram based conventional metrics. 

We present two examples in Table \ref{table:tab_caps}. It can be seen clearly that the captions generated by the MLE model differ only slightly to each other at the end of the captions. In contrast, the captions generated by the C-GAN model are more diverse. For example, in example \num{1}, `buzzing' and `drilling' are used to describe the sound while `jackhammer' and `engine' describes the objects which make sound. In example \num{2}, different verbs are used to describe the `rain', where some also describe the `wind', while others don't. 

\subsection{Ablation studies}
\label{ssec:ablation}
We conduct experiments to study the contribution of discriminator and evaluators in our proposed framework. The results are shown in Table \ref{table:ablation_results}. When only using the naturalness discriminator, the generated captions can be very diverse but get low scores on the $n$-gram based evaluation metrics. This is because the model would generate random captions which are irrelevant to the given audio clip, thus these captions would be diverse but were not semantically faithful to the audio clip. When only using the evaluators, the proposed framework degenerates to a conventional RL optimization method. When only using the semantic evaluator, the performance of the system drops significantly and get scores near \num{0} for the $n$-gram based evaluation metrics. This is because the system cannot even generate reasonable sentences, but instead some disordered and repetitive words. When only optimizing CIDE$_r$, the $n$-gram based evaluation metrics drops slightly, compared to the MLE model, as a random vector is appended to encourage the diversity of generated captions. The mBLEU$_4$ is close to that of the MLE model using beam search, but the vocabulary size and div-$n$ is small which are consistent with the observation in \cite{mei2021encoder} that optimizing CIDE$_r$ using RL would reduce distinctiveness of the generated captions. Overall, the results suggest that the discriminator and semantic evaluator are complementary to ensure the accuracy and diversity of the captions while the language evaluator can improve the metrics measuring accuracy but may impact adversely on the diversity in the generated captions. 

\section{CONCLUSION}
\label{sec:Conclusion}

This paper proposed an alternative approach to audio captioning using conditional generative adversarial network (C-GAN). We focused on improving the diversity of generated captions which was neglected in the literature. The proposed framework is composed of a generator, a discriminator and two fixed evaluators. The generator and discriminator are trained alternatively during training while two evaluators are used to evaluate the generated captions and give guidance to the generator with desired properties. We empirically show that the system trained via the proposed framework can generate more diverse captions although the accuracy scores in terms of conventional evaluation metrics may drop. Future research should be carried out to improve the quality of the captions as well as preserving the diversity.

\section{ACKNOWLEDGMENT}
\label{sec:ack}
This work is partly supported by grant EP/T019751/1 from the Engineering and Physical Sciences Research Council (EPSRC), a Newton Institutional Links Award from the British Council, titled “Automated Captioning of Image and Audio for Visually and Hearing Impaired” (Grant number 623805725) and a Research Scholarship from the China Scholarship Council (CSC) No. 202006470010.

\bibliographystyle{IEEEtran}
\bibliography{strings,refs}

\end{document}